\documentstyle[aps,epsf,twocolumn]{revtex} 
\tightenlines 
\begin{document} 
\draft 
\title{
Magnon Exchange Mechanism of Superconductivity: $ZrZn_2$, $URhGe$  }
\author{Naoum Karchev\cite{byline}}
\address{
Department of Physics, University of Sofia, 1126 Sofia, Bulgaria
}
%      
%\date (Received \today)
\maketitle

\begin{abstract}

The magnon exchange mechanism of superconductivity was developed to explain in
a natural way the fact that the 
superconductivity in $UGe_2$, $ZrZn_2$ and 
$URhGe$ is confined to the ferromagnetic 
phase.The order parameter is a spin anti-parallel component of 
a spin-1 triplet with zero spin projection. The transverse spin 
fluctuations are pair forming and the longitudinal ones are pair 
breaking. In the present paper, a superconducting solution, based on the
magnon exchange mechanism, is obtained which closely matches the experiments
with $ZrZn_2$ and $URhGe$. The onset of superconductivity leads to the appearance of
complicated Fermi surfaces in the spin up and spin down momentum 
distribution functions. Each of them consist of two pieces, but they are
simple-connected and can be made very small by varying the microscopic parameters.
As a result, it is obtained that the specific heat depends on the temperature
linearly, at low temperature, and the coefficient $\gamma=\frac {C}{T}$ is
smaller in the superconducting phase than in the ferromagnetic one.
The absence of a quantum transition from ferromagnetism to ferromagnetic
superconductivity in a weak ferromagnets $ZrZn_2$ and $URhGe$ is explained accounting for
the contribution of magnon self-interaction to the spin fluctuations' parameters. It is
shown that in the presence of an external magnetic field the system undergoes
a first order quantum phase transition. 
  
\end{abstract}


Very recently ferromagnetic superconductivity (f-superconductivity) 
has been observed 
in $UGe_2$\cite{me1}, $ZrZn_2$\cite{me2} and $URhGe$\cite{me3}.
The superconductivity is confined to the ferromagnetic phase. 
Ferromagnetism and superconductivity are believed to arise due to the same 
band electrons. The persistence of ferromagnetic order
within the superconducting phase has been ascertained by neutron scattering.
The specific heat anomaly associated with the superconducting transition 
in these materials appears to be absent.

At ambient pressure $UGe_2$ is an itinerant ferromagnet below the Curie
temperature
 $T_c=52K$, with low-temperature ordered moment of $\mu_s=1.4\mu_B/U$. 
With increasing pressure the system passes through two
successive quantum phase transition, from ferromagnetism to
f-superconductivity at $P\sim 10$ kbar, and at higher pressure $P_c\sim$ 16
kbar to paramagnetism\cite{me1,me4}.At the pressure where the superconducting
transition temperature is a maximum $T_{sc}=0.8K$,
the ferromagnetic state is still
 stable with $T_c=32K$, and
an ordered moment about 
 $1.0\mu_B/U$\cite{me1,me4,me5}. The specific heat coefficient 
$\gamma=\frac {C}{T}$ increases
steeply near 11 kbar and retains a large and nearly 
constant value\cite{me5}.  

The ferromagnets $ZrZn_2$ and $URhGe$ are superconducting at ambient pressure 
with superconducting critical temperatures $T_{sc}=0.29K$ and $T_{sc}=0.25K$ 
respectively.
$ZrZn_2$ is ferromagnetic below the Curie temperature $T_c=28.5K$ with 
low-temperature ordered moment of $\mu_s=0.17\mu_B$ per formula unit, while 
for $URhGe$\,\,
$T_c=9.5K$
and $\mu_s=0.42\mu_B$. The low Curie temperatures and small ordered moments
indicate that compounds are close to a ferromagnetic quantum critical point.
A large jump in the specific heat, at the temperature where the resistivity
becomes zero, is observed in $URhGe$. At low temperature the specific
heat coefficient $\gamma$ is twice smaller than in the ferromagnetic phase.

The most popular theory of f-superconductivity is based on the paramagnon
exchange mechanism\cite{me6,me7}. The order parameters are spin
parallel components of the spin triplet. The superconductivity in
$ZrZn_2$ was predicted, but the theory meets many difficulties. 
In order to explain the absence of superconductivity in paramagnetic phase 
it was accounted for the magnon
paramagnon interaction and proved that the critical temperature is much
higher in the ferromagnetic phase than in the paramagnetic one\cite{me8}.
To the same purpose, the Ginzburg-Landau mean-field theory was modified
with an exchange-type interaction between the magnetic moments of 
triplet-state Cooper pairs and the ferromagnetic 
magnetization density\cite{me9}.In \cite{me10} the authors make the 
important assumption that
only majority spin fermions form pairs.Then, only minority
spin fermions contribute to the asymptotic of the specific heat, and the
coefficient $\gamma=\frac {C}{T}$ is twice smaller in the superconducting
phase.The result closely matches the experiments with $URhGe$\cite{me3},
but does not resemble the experimental results for $UGe_2$ and $ZrZn_2$.
The assumption seems to be doubtful for systems with very small 
ordered moment. 
Despite of the efforts, the improved theory of paramagnon induced
superconductivity can not cover the whole variety of properties of 
f-superconductivity. 

In the present paper an itinerant system is considered in which the
spin-$\frac 12$ fermions $c_{\sigma}(\vec x)(c^+_{\sigma}(\vec x))$  
responsible for the ferromagnetism are the same
quasiparticles which form the Cooper pairs. The exchange of 
spin fluctuations leads to an effective four fermion theory. It describes  
the
interaction of the components of spin-1 composite fields $(\uparrow\uparrow,
\uparrow\downarrow+\downarrow\uparrow, \downarrow\downarrow)$ which have a
projection of spin 1,0 and -1 respectively, and the
interaction of the spin singlet composite fields
$\uparrow\downarrow-\downarrow\uparrow$. The spin singlet fields' interaction
is repulsive and does not contribute to the superconductivity \cite{me11}. The
spin parallel fields' interactions are due to the exchange of paramagnons and
do not contribute to the magnon-mediated superconductivity. The relevant
interaction is that of the $\uparrow\downarrow+\downarrow\uparrow$ fields. 
The potential of this interaction has an attracting part due to exchange 
of magnons and a repulsive part due to exchange of paramagnons.  

By means of the Hubbard-Stratanovich transformation one introduces 
$\uparrow\downarrow+\downarrow\uparrow$ composite field and then the fermions
can be integrated out. The obtained free energy is a function
of the composite field and the integral over the composite field can be
performed approximately by means of the steepest descend method. To this end
one sets the first derivative of the free energy with respect to composite
field equal to zero, this is the gap equation, and looks for a solution which
minimizes the free energy.

The gap is an antisymmetric function $\Delta (-\vec k)=-\Delta (\vec k)$, so 
that the expansion in terms of spherical harmonics $Y_{lm}(\Omega_{\vec k})$ 
contains only terms with odd $l$. I assume that the component with $l=1$ and 
$m=0$ is nonzero and the other ones are zero  
\FL 
\begin{equation} 
\Delta (\vec k)=\Delta_{10}(k)\sqrt {\frac {3}{4\pi}}\cos\theta. 
\label{me1} 
\end{equation} 
Expending 
the potential in terms of Legendre polynomial $P_l$ one obtains that only the 
component with $l=1$ contributes the gap equation. The potential $V_1(p,k)$ 
has the form, 
\FL 
\begin{eqnarray} 
V_{1}(p,k) & = & \frac {3M}{\rho}\left[\frac 
{p^2+k^2}{4p^2k^2}\ln\left(\frac {p+k}{p-k}\right)^2-\frac{1}{pk}\right] 
- \nonumber \\ 
& &  
\frac {3M}{\rho}\beta \left[\frac {p^2+k^2}{4p^2k^2}\ln\frac 
{r'+(p+k)^2}{r'+(p-k)^2}\,-\,\frac {1}{pk}\right],  
\label{me2} 
\end{eqnarray} 
where $M$ is zero 
temperature dimensionless magnetization of the system per lattice site and 
$\rho$ is the spin stiffness 
constant which is proportional to $M$ ($\rho=M\rho_0$) The constants 
$\beta,\rho_0$ and $b$ are phenomenological ones subject to the relation 
$\beta=\frac {\rho}{2Mb}=\frac {\rho_0}{2b}>1$, and 
$r'=\frac {r}{b}<<1$, where the 
parameter $r$ is the inverse static longitudinal magnetic susceptibility, 
which measures the deviation from quantum critical point. 
A straightforward analysis  
shows that for a fixed $p$ , the potential is positive when $k$ 
runs an interval around $p$ $(p-\Lambda,p+\Lambda)$, 
where $\Lambda$ is approximately independent on $p$.  
In order to allow for an explicit analytic solution, I introduce further 
simplifying assumptions by neglecting the dependence of $\Delta_{10}(k)$ 
on $k$ ($\Delta_{10}(k)=\Delta_{10}(p_f)=\Delta, p_{f}=\sqrt {2\mu m} $) 
and setting 
$V_1(p_f,k)$ equal to a constant $V_1$ within interval 
$(p_f-\Lambda,p_f+\Lambda)$ and to zero elsewhere.

To ensure that the
fermions which form Cooper pairs are the same as those responsible for
spontaneous magnetization, one has to consider the equation for the
magnetization
\FL
\begin{equation} 
M=\frac 12 <c^+_{\uparrow}c_{\uparrow}-c^+_{\downarrow}c_{\downarrow}>
\label{me3}
\end{equation}
as well.
Then the system of equations for the gap and for the magnetization determines
the phase where the superconductivity and the ferromagnetism coexist.The system 
can be written in terms of Bogoliubov excitations, which have
the following dispersions relations:
\FL
\begin{eqnarray}
& & E_1(\vec k) = -\frac
{JM}{2}-\sqrt{\epsilon^2(\vec k)+|\Delta(\vec
k)|^2} \nonumber \\ 
& & E_2(\vec k) = \frac
{JM}{2}-\sqrt{\epsilon^2(\vec
k)+|\Delta(\vec k)|^2}  
\label{me4}
\end{eqnarray}
where $\Delta(\vec k)$
is the gap (\ref{me1}),
$J$ is the spin exchange constant, 
and $\epsilon(\vec k)=\frac {\vec k^2}{2m}-\mu$.

At zero temperature the equations take the form
\FL
\begin{eqnarray}
M & = & \frac
{1}{8\pi^2}\int\limits_0^{\infty}dkk^2\int\limits_{-1}^{1}dt[1-\Theta(-E_2(k,t))]
\label{me5}\\
\Delta & = & \frac {J^2V_1}{32\pi^2}\int\limits_{p_f-\Lambda}^{p_f+\Lambda} dk
k^2\int\limits_{-1}^{1}dt\,t^2\frac
{\Theta(-E_2(k,t))}{\sqrt{\epsilon^2(k)+\frac {3}{4\pi}t^2\Delta^2}}
\Delta
\label{me6}
\end{eqnarray}
where $t=\cos\theta$.

The solution of the system which satisfies $\sqrt {\frac
{3}{\pi}}\Delta<JM$ is discussed in \cite{me12}. In the present paper one
looks for a solution of the system which satisfies
\FL
\begin{equation}
\sqrt {\frac {3}{\pi}}\Delta\,>\,JM
\label{me7}
\end{equation} 

One is primarily interested in determining at what 
magnetization a superconductivity exists. The inequality Eq.(\ref{me7})
shows that the gap can not be arbitrarily small when the magnetization
is finite. Hence the system undergoes the quantum phase transition from
ferromagnetism to f-superconductivity with a jump. Approaching the quantum
critical point from the ferromagnetic side, one sets the gap equal to zero
in the equation for the magnetization (\ref{me5}) and considers the gap 
equation (\ref{me6}) with magnetization as a parameter. It is more 
convenient to consider the free energy as a function of the gap for the
different values of the parameter $M$. To this purpose I introduce the
dimensionless "gap" $x$ and the parameters $s,\lambda$ and $g$ 
\FL
\begin{equation}
x=\sqrt {\frac {3}{\pi}}\frac {m}{p_{f}^2}\Delta,\,\,
s=\frac {m}{p_{f}^2}JM,\,\,\lambda=\frac {\Lambda}{p_{f}},\,\,
g=\frac {J^2V_1mp_{f}}{8\pi^2} 
\label{me8}
\end{equation}
Then the free energy is a function of $x$
and depends on the parameters $s, \lambda$ and $g$.
\FL
\begin{eqnarray}
& & F(x)=\frac {6m^2}{\pi p_{f}^4}\left({\cal F}(x)-{\cal F}(0)\right) = x^2+ 
g\int\limits_{1-\lambda}^{1+\lambda}dq q^2\int\limits_{-1}^1dt\,\times\nonumber
\\ & & \left[\left(s-\sqrt {(q^2-1)^2+t^2x^2}\right)
\Theta(\sqrt{(q^2-1)^2+t^2x^2}-s)-\right. \nonumber \\
& & \left. \left(s-\sqrt {(q^2-1)^2}\right)
\Theta(\sqrt{(q^2-1)^2}-s)\right]
\label{me9}
\end{eqnarray}
The dimensionless free energy $F(x)$ is depicted in Fig.1 for
$\lambda=0.08,\,g=20$ and three values of the parameter $s$, $s=0.8, s=0.69$
and $s_{cr}=0.595$. As the graph shows, for some values of the microscopic
parameters $\lambda$ and $g$, and decreasing the parameter $s$ (the
magnetization), the system passes trough a first order quantum phase
transition. The critical values $s_{cr}$ and $x_{cr}$ satisfy $\frac
{x_{cr}}{s_{cr}}=\sqrt {\frac {3}{\pi}}\frac {\Delta_{cr}}{JM_{cr}}>1$ in
agreement with Eq.(\ref{me7}).
\begin{figure}[h]  
\vspace{0.0cm}  
\epsfxsize=8.0cm  
\hspace*{-0.15cm}  
\epsfbox{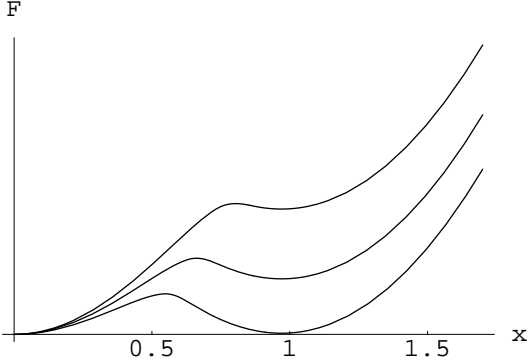}   
\caption{The dimensionless free energy $F(x)$ as a function of
dimensionless gap $x$. $\lambda=0.08,\,g=20$, $s_1=0.8$(upper line),
$s_2=0.69$(middle line) and
 $s_{cr}=0.595$(lower line).}   
\label{fig1}  
\end{figure}  

Varying the microscopic parameters beyond the critical values, one has to solve
the system of equations (\ref{me5},\ref{me6}).The equation of magnetization
(\ref{me5}) shows that it is convenient to represent the gap in the form 
$\Delta= \sqrt {\frac {\pi}{3}}\kappa (M) JM$, where $\kappa (M)>1$.
Then the equation $E_2(k,t)=0$, which defines the Fermi surface, has no
solution if   $-1<t<-\frac {1}{\kappa (M)}$ and $\frac {1}{\kappa(M)}<t<1$,
and has two solutions
\FL 
\begin{equation}
p^{\pm}_{f}=\sqrt {p^2_{f}\pm m\sqrt{J^2M^2-\frac
{3}{\pi}t^2\Delta^2}}
\label{me10}
\end{equation} 
when $-\frac {1}{\kappa(M)}<t<\frac {1}{\kappa (M)}$. The solutions
(\ref{me10}) determine the two pieces of the Fermi surface. They stick
together at $t=\pm\frac {1}{\kappa(M)}$, so that the Fermi surface is simple
connected. The domain between pieces contributes to the
magnetization $M$ in
 Eq.(\ref{me5}), but it is cut out from the domain of
integration in the gap
 equation Eq.(\ref{me6}). 

When the magnetization approaches
zero, one can approximate the
equation for magnetization Eq.(\ref{me5}) substituting $p^{\pm}_{f}$ from
Eq.(\ref{me10}) in the
the difference $(p^{+}_{f})^2-(p^{-}_{f})^2$ and setting
$p^{\pm}_{f}=p_{f}$ elsewhere. Then, in this approximation, the
magnetization is linear in $\Delta$, namely
\FL
\begin{equation}
\Delta =\sqrt {\frac {\pi}{3}}J\kappa M
\label{me11}
\end{equation}
where $\kappa=\frac {mp_{f}J}{16\pi}$ is the small magnetization limit of
$\kappa(M)$.
The Eq.(\ref{me11}) is a solution if $mp_{f}J>16\pi$ (see Eq.(\ref{me7})).
Substituting $M$ from
 Eq.(\ref{me11}) in Eq.(\ref{me6}), one arrives at an
equation for the gap. This
 equation can be solved in a standard way and the
solution is
 
\FL
\begin{equation}
\Delta\,=\,\sqrt {\frac {16\pi}{3}}
\frac {p_{f}\Lambda}{m}
\exp \left[-\frac {24\pi^2}{mp_{f}J^2V_1}-\frac {\pi}{4\kappa^3}+\frac
{1}{3}\right]
\label{me12}
\end{equation}
Eqs (\ref{me11},\ref{me12}) are the solution of the system
Eqs.(\ref{me5},\ref{me6}) near the quantum transition to paramagnetism. 
The second derivative of the free energy Eq.(\ref{me9}) with respect
to the gap is positive when $\frac {mp_{f}J}{16\pi}>(\frac {21\pi}{16})^{\frac
13}$,  hence the state where the
superconductivity and the ferromagnetism coexist is stable.

When superconductivity and
ferromagnetism coexist, the momentum
 distribution functions
$n^{\uparrow}(p,t)$ and $n^{\downarrow}(p,t)$ of the
 spin-up and spin-down
quasiparticles have complicated Fermi surfaces. One can
 write them in
terms of the distribution functions of the Bogoliubov fermions
 
\FL
\begin{eqnarray}
n^{\uparrow}(p,t) & = & u^2(p,t)n_1(p,t)+v^2(p,t)n_2(p,t)
\label{me13} \\
n^{\downarrow}(p,t) & = & u^2(p,t)(1-n_1(p,t))+v^2(p,t)(1-n_2(p,t))
\nonumber
\end{eqnarray}
where $u(p,t)$ and $v(p,t)$ are the coefficients in the Bogoliubov
transformation. At zero temperature $n_1(p,t)=1$,
$n_2(p,t)=\Theta(-E_2(p,t))$, and the Fermi surface Eq.(\ref{me10}) manifests
itself both in the spin up and spin-down momentum distribution functions.
The functions are depicted in Fig.2 and Fig.3.
 
\begin{figure}[h] 
\vspace{0.3cm} 
\epsfxsize=7.0cm 
\hspace*{0.2cm} 
\epsfbox{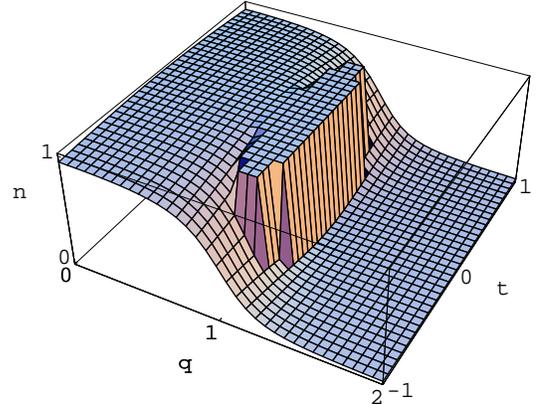} 
%\vspace{-5.2cm} 
\caption{The zero temperature momentum distribution $n$, for spin up 
fermions, as a function of $q=\frac {p}{p_{f}}$ and $t=\cos\theta$.} 
%\vspace{-0.26cm} 
\label{fig2} 
\end{figure} 

\begin{figure}[h] 
\vspace{0.3cm} 
\epsfxsize=7.0cm 
\hspace*{0.2cm} 
\epsfbox{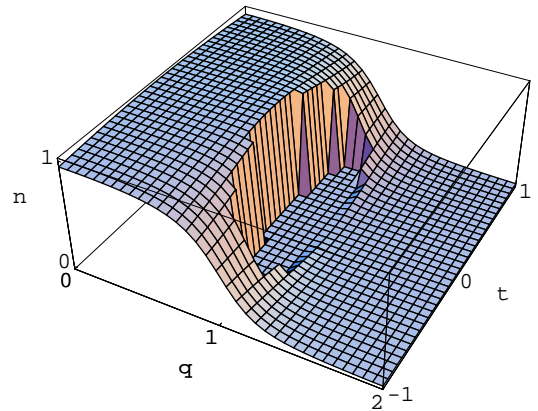} 
%\vspace{-5.2cm} 
\caption{The zero temperature momentum distribution $n$, for spin down
fermions, as a function of $q=\frac {p}{p_{f}}$ and $t=\cos\theta$.} 
%\vspace{-0.26cm} 
\label{fig3} 
\end{figure}

The existence of the Fermi surface explains the linear dependence
of the specific heat at low temperature:
\FL
\begin{equation}
\frac {C}{T}\,=\,\frac {2\pi^2}{3}N(0)
\label{me14}
\end{equation}
Here $N(0)$ is the density of states on the Fermi surface.
One can rewrite the $\gamma=\frac {C}{T}$ constant in terms of Elliptic
Integral of the second kind $E(\alpha,x)$
\FL
\begin{eqnarray}
\gamma\,=\,\frac {m p_{f}}{3\kappa(M)} & &
\left[(1+s)^{\frac 12} E(\frac {\pi}{4},\frac
{2s}{s+1})+\right. \nonumber  \\  
& & \left. (1-s)^{\frac 12} E(\frac {\pi}{4},\frac
{2s}{s-1})\right].
\label{me15}
\end{eqnarray}
where $s<1$ (see Eq.(\ref{me8})).
Eq.(\ref{me15}) shows that for $\kappa(M)>>1$ the specific heat
constant $\gamma$  is small in f-superconducting phase, which
closely matches the
 experiments with $ZrZn_2$ and $URhGe$.

An important experimental fact is that $ZrZn_2$ and $URhGe$ are
superconductors at ambient pressure as opposed to the existence of a quantum
phase transition in $UGe_2$. To comprehend this difference one
considers the potential (\ref{me2}). The quantum phase transition results from
the existence of a momentum cutoff $\Lambda$, above which the potential is
repulsive. In turn, the cutoff excistence follows from the relation
$\beta=\frac {\rho}{2Mb}>1$, which is true when the spin-wave approximation
expression for the spin stiffness constant $\rho=M\rho_0$ is used. The spin
wave approximation correctly describes systems with a large magnetization, for
example $UGe_2$. But in order to study systems with small magnetization, one
has to
account for the magnon-magnon interaction which changes the small
magnetization  asymptotic of $\rho$, $\rho=M^{1+\alpha}\rho_0$, where
$\alpha>0$. Then for a small $M$ $\beta<1$, and the potential is attractive
for
 all momenta. Hence for systems which, at ambient pressure, are close to
quantum
 critical point, as $ZrZn_2$ and $URhGe$, the magnon self-interaction
renormalizes
 the spin fluctuations parameters so that the magnons dominate
the pair
 formation and quantum phase transition can not be observed. But if
one applies
 an external magnetic field, the magnon opens a gap proportional
to the
 magnetic field. Increasing the magnetic field the paramagnon
domination leads
 to first order quantum phase transition.  

The proposed model of ferromagnetic superconductivity differs from the
models discussed in \cite{me6,me7,me8,me9,me10} in many aspects. First, the
superconductivity is due to the exchange of magnons, and the model describes
in an unified way the superconductivity in $UGe_2,\,ZrZn_2$ and $URhGe$.
Second, the solution Eq.(\ref{me11})shows that magnetization and
superconductivity disappear simultaneously. It results from the equation of
magnetization, which in turn is added to ensure that the 
fermions which form Cooper pairs are the same as those responsible for 
spontaneous magnetization. Hence, the fundamental assumption that
superconductivity and ferromagnetism are caused by the same electrons leads
to the experimentally observable fact that the quantum phase transition is a
transition to paramagnetic phase without superconductivity.
Third, the paramagnons have
pair-breaking effect. So, the understanding the mechanism of paramagnon
suppression is crucial in the search for the ferromagnetic superconductivity
with higher critical temperature. For example, one can build such a bilayer
compound that the spins in the two layers are oriented in two
non-collinear directions, and the net ferromagnetic moment is nonzero.
The paramagnon in this phase is totally suppressed and the low lying excitations
consist of magnons and additional spin wave modes with linear dispersion
$\epsilon(k)\sim k$\cite{me13}. If the new spin-waves are pair breaking, their
effect is weaker than those of the paramagnons, and hence the superconducting
critical temperature should be higher.


\begin{references}
%
\frenchspacing
%
\bibitem[*]{byline} Electronic address: naoum@phys.uni-sofia.bg
% 
\bibitem{me1} S. Saxena, P. Agarwal, K. Ahilan, F. M. Grosche, R. Haselwimmer,
M. Steiner, E. Pugh, I. Walker, S. Julian, P. Monthoux, G. Lonzarich, A.
Huxley, I. Sheikin, D. Braithwaite, and J. Flouquet, 
Nature (London) {\bf406}, 587 (2000).
%
\bibitem{me2} C. Pfleiderer, M. Uhlarz, S. Hayden, R. Vollmer, H.v.
L\"ohneysen, N. Bernhoeft, and G. Lonzarich, 
Nature (London) {\bf 412}, 58
(2001).
%
\bibitem{me3} D. Aoki, A. Huxley, E. Ressouche, D. Braithwaite, J. Flouquet,
J-P. Brison, E.Lhotel, and C. Paulsen, 
Nature (London) {\bf 413}, 613 (2001).
%
\bibitem{me4} A. Huxley, I. Sheikin, E. Ressouche, N. Kernavanois, D.
Braithwaite, R. Calemczuk, and J. Flouquet, 
Phys. Rev. B {\bf 63}, 144519
(2001).
%
\bibitem{me5} N. Tateiva, T. Kobayashi, K. Hanazono, K. Amaya, Y. Haga, R.
Settai, and Y. Onuki, 
J. Phys. Condens. Matter {\bf
13}, L17 (2001).
%
\bibitem{me6} C. P. Enz, and P. T. Matthias,  
Science {\bf 201}, 828 (1978). 
% 
\bibitem{me7} D. Fay, and J. Apple,  
Pys.Rev. B {\bf 22}, 3173 
(1980). 
%
\bibitem{me8}
T. Kirkpatrick, D. Belitz, T. Vojta and R. Narayanan
Phys. Rev. Lett. {\bf 87}, 127003 (2001).
%
\bibitem{me9} M. B. Walker, and K. V. Samokhin, Phys. Rev. Lett. {\bf 88},
207001 (2002).  
%
\bibitem{me10} K. Machida, and T. Ohmi, 
Phys. Rev. Lett.{\bf 86}, 850 (2001).
%
\bibitem{me11} N. F. Berk, and J. R Schrieffer, 
Phys. Rev. Lett. {\bf 17}, 433 (1966).
%
\bibitem{me12} N. Karchev, cond-mat/0206534 (2002)
%
\bibitem{me13} S. Sachdev, and T. Senthila,
Ann.Phys.(NY) {\bf 251}, 76 (1996)
%

\end{references}
\end{document}